\documentclass[conference]{IEEEtran}
\IEEEoverridecommandlockouts

\usepackage{cite}
\usepackage{amsmath,amssymb,amsfonts,mathtools}
\usepackage{algorithmic}
\usepackage{algorithm}
\usepackage{graphicx}
\usepackage{textcomp}
\usepackage{xcolor}
\def\BibTeX{{\rm B\kern-.05em{\sc i\kern-.025em b}\kern-.08em
    T\kern-.1667em\lower.7ex\hbox{E}\kern-.125emX}}
\begin{document}

\newtheorem{thm}{Theorem}[section]
\newtheorem{prop}[thm]{Proposition}
\newtheorem{cor}[thm]{Corollary}
\newtheorem{lem}[thm]{Lemma}
\newtheorem{conj}[thm]{Conjecture}

\allowdisplaybreaks

\renewcommand{\algorithmicrequire}{\textbf{Input:}}
\renewcommand{\algorithmicensure}{\textbf{Output:}}

\newcommand{\E}{\mathbb{E}}
\newcommand{\F}{\mathbb{F}}
\newcommand{\N}{\mathbb{N}}
\newcommand{\R}{\mathbb{R}}
\newcommand{\dist}{\mathrm{dist}}
\newcommand{\vbl}{\mathrm{vbl}}

\title{Asymptotic Rate Bounds and Constructions\\
for the Inclusive Variant of Disjunct Matrices\\
\thanks{This work was supported by JST SPRING Grant Number JPMJSP2109 (Y.M.) and JSPS KAKENHI Grant Number JP24K14815 (Y.F.).}
}

\author{\IEEEauthorblockN{Yuto Mizunuma}
\IEEEauthorblockA{Graduate School of Informatics \\
Chiba University\\
1-33 Yayoi-Cho Inage-Ku, Chiba 263-8522, Japan \\
Email: mizunuma@chiba-u.jp}
\and
\IEEEauthorblockN{Yuichiro Fujiwara}
\IEEEauthorblockA{Faculty of Informatics \\
Chiba University\\
1-33 Yayoi-Cho Inage-Ku, Chiba 263-8522, Japan \\
Email: yuichiro.fujiwara@chiba-u.jp}
}

\maketitle

\begin{abstract}
Disjunct matrices, also known as cover-free families and superimposed codes, are combinatorial arrays widely used in group testing.
Among their variants, those that satisfy an additional combinatorial property called inclusiveness form a special class suitable for computationally efficient and highly error-tolerant group testing under the general inhibitor complex model, a broad framework that subsumes practical settings such as DNA screening.
Despite this relevance, the asymptotic behavior of the inclusive variant of disjunct matrices has remained largely unexplored.
In particular, it was not previously known whether this variant can achieve an asymptotically positive rate, a requirement for scalable group testing designs.
In this work, we establish the first nontrivial asymptotic lower bound on the maximum achievable rate of the inclusive variant, which matches the strongest known upper bound up to a logarithmic factor.
Our proof is based on the probabilistic method and yields a simple and efficient randomized construction.
Furthermore, we derandomize this construction to obtain a deterministic polynomial-time construction.
These results clarify the asymptotic potential of robust and scalable group testing under the general inhibitor complex model.
\end{abstract}


\section{Introduction}
Group testing is a framework for efficiently identifying a small set of target items, called \emph{defectives}, within a large population.
The key idea is to place items into a small number of pools and test the pools rather than testing items individually.
Introduced by Dorfman in 1943 for economical large-scale
blood screening of military conscripts for syphilis~\cite{Dor43}, group testing has since found applications in diverse areas, such as biology~\cite{FKKM97,DH00}, neuroscience~\cite{BKCE19}, drug discovery~\cite{XTSY01}, networking~\cite{DPSV19}, compressed sensing~\cite{AS12}, and more recently COVID-19 screening~\cite{SLWSSOGSEGGMSFNSPH20}.

This paper focuses on \emph{nonadaptive} group testing~\cite{DH00,DH06}, in which, as opposed to its adaptive counterpart~\cite{Sca18,AJS19}, the pools are fixed in advance and cannot adapt to previous test outcomes.
Nonadaptive group testing is particularly useful when testing each pool takes a long time but can be done in parallel.

In nonadaptive group testing, special combinatorial arrays called \emph{disjunct matrices} are widely used for efficient and robust pooling designs~\cite{DH06}.
In fact, in the classical setting, a suitable $t \times n$ disjunct matrix allows us to design $t$ pools and infer the locations of all defectives among $n$ items in time $O(t n)$ from the $t$ test outcomes, even when some test outcomes are erroneous~\cite{Mac97,DH06,CFH08}.

A central objective in group testing is to handle a large population of items with a small number of tests.
Ideally, the number of tests should be logarithmic in the population size.
Hence, borrowing terminology from coding theory, we consider the \emph{rate} $\frac{\log_2 n}{t}$ of a $t \times n$ disjunct matrix as a key parameter for testing efficiency and hope to construct disjunct matrices with large rates.
As described in Section~\ref{sec:inclusively_disjunct_matrices}, conventional disjunct matrices can achieve asymptotically positive rates~\cite{SW04,IM24}, which implies that we can identify all defectives by conducting only a logarithmic number of tests in the population size in the classical group testing model.

In the literature, various models of group testing have been proposed~\cite{DH00,DH06}.
The classical group testing model assumes that each positive outcome is triggered by a single item and that, in the absence of testing errors, a pool tests positive if and only if it contains at least one defective item.
However, many practical applications cannot be fully captured within this classical framework.
For instance, in applications that focus on combinations of items with specific properties, one aims to identify targets among a collection of \emph{complexes}, i.e., nonempty subsets of the population, rather than among individual items~\cite{LH04,LTLW05,CDH07}.
This is the motivation for generalizing the classical group testing model to the \emph{complex model}~\cite{Tor99}.

Another example that the classical setting cannot handle is the presence of troublesome items or complexes called \emph{inhibitors}, whose presence in a pool can potentially mask or suppress the positive outcome that defectives would otherwise cause~\cite[Ch.~8]{DH06}.
The contexts in which phenomena of this kind have been reported include blood testing~\cite{PS94}, drug discovery~\cite{XTSY01}, and DNA library screening~\cite{BLC91}.

Inhibitors in group testing were first systematically studied in~\cite{FKKM97}, which introduced a simple model for the negative effects of certain molecular biological substances on group testing.
Among many approaches to group testing involving inhibitors, a particularly general framework, called the \emph{general inhibitor model}, was introduced in~\cite{HC07}, which assumes the possible presence of inhibitors but provides no further exploitable information about when or how they cause false negatives.
Subsequently, \cite{CCH11} proposed the \emph{general inhibitor complex model}, combining the complex model and the general inhibitor model.
In this model, both defectives and inhibitors are treated as types of complexes, and as in the general inhibitor model, the specific behavior of inhibitors is not assumed.
Since the general inhibitor complex model can handle a broad range of practical applications, studying efficient inference methods and theoretical limitations for this model is highly valuable.

Because inhibitors in this general setting may behave in any adversarial way, it is quite challenging to design effective group testing schemes.
Nevertheless, in principle, since false negatives are still errors, one could address adversarial inhibitors using highly error-tolerant schemes designed for group testing without inhibitors.
Unfortunately, this approach has so far led only to computationally very expensive inference algorithms.
Indeed, a $t \times n$ conventional error-tolerant disjunct matrix requires time $O(t n^h)$ to determine whether each complex is defective, assuming that there are at most $h$ inhibitors among the complexes~\cite{CCH11}.

To reduce this time complexity, \cite{CCF10} introduced an additional property of disjunct matrices, called \emph{inclusiveness}, which guarantees that each defective can be masked by inhibitors in only a limited number of pools.
By using the inclusive variant of disjunct matrices, which we refer to as \emph{inclusively disjunct matrices}, the time to determine whether each complex is defective can be reduced to $O(t)$, which is far more practical than the $O(t n^h)$-time scheme.
Therefore, to cope with inhibitors of unknown behavior efficiently, it suffices to design inclusively disjunct matrices with large rates.

Despite this relevance, the asymptotic behavior of the inclusive variant has remained largely unexplored.
In particular, although applications typically require large positive rates, it was not previously known whether the class of inclusively disjunct matrices can achieve asymptotically positive rates.

In this work, we establish the first nontrivial asymptotic lower bound on the maximum achievable rate of inclusively disjunct matrices, which matches the strongest known upper bound up to a logarithmic factor.
Our proof is based on the probabilistic method~\cite{AS16} and yields a simple and efficient randomized construction.
Furthermore, we derandomize this construction via the method of conditional expectations~\cite{MU17} to obtain a deterministic polynomial-time construction.
These results show that the class of inclusively disjunct matrices admits asymptotically positive rates with efficient constructions and clarify the asymptotic potential of robust and scalable group testing under the general inhibitor complex model.

In the next section, we formally define inclusively disjunct matrices and review known results.
We prove the first nontrivial asymptotic lower bound on the maximum achievable rate of inclusively disjunct matrices and give an efficient randomized construction in Section~\ref{sec:lower_bound_randomized_construction}, and then derandomize it in Section~\ref{sec:derandomized_construction}.
Finally, Section~\ref{sec:concluding_remarks} concludes this paper with some remarks.

\section{Inclusively disjunct matrices}
\label{sec:inclusively_disjunct_matrices}
Here, we formally define inclusively disjunct matrices and give a brief review of known results.
For a more thorough treatment, including basic theory of nonadaptive group testing and background from bioinformatics, we refer the reader to~\cite{DH00,DH06,IM24} and references therein.

We begin by formally defining disjunct matrices.
For a positive integer $m$, let $[m]$ denote the set $\{ 1, 2, \dots, m \}$.
Let $T = (t_{ij})$ be a $t \times n$ binary matrix over $\{ 0, 1 \}$.
For each $j \in [n]$, let $C_j \coloneqq \{ i \in [t] : t_{ij} = 1 \}$ be the support of the $j$th column vector.
For disjoint sets $A, B \subseteq [n]$, define
\begin{equation*}
  Z_T(A, B) \coloneqq \left| \left( \bigcap_{j \in B} C_{j} \right) \setminus \left( \bigcup_{j \in A} C_{j} \right) \right|
\end{equation*}
to be the number of rows in which all the $|A|$ entries in the columns indexed by the elements of $A$ are $0$ and all the $|B|$ entries in the columns indexed by the elements of $B$ are $1$.
For positive integers $d$, $r$, and $z$, the matrix $T$ is \emph{$(d, r; z]$-disjunct} if for any pair of disjoint sets $A, B \subseteq [n]$ with $|A| = d$ and $|B| = r$, $Z_T(A, B) \geq z$.

Disjunct matrices correspond to the most popular and fundamental group testing schemes used in various situations~\cite{Mac97,DH00,DH06,CCF10,AJS19,BCE21}.
They are also known as superimposed codes~\cite{KS64,BMTW84,Wol85,DMR00,BV03,FC10,GRV20,RV23} in coding theory as well as cover-free families~\cite{EFF82,EFF85,Rus94,DVTM02,Yeh02,SW04,DSW04,IM24} in extremal combinatorics.
For their applications outside group testing and known existence results, we refer the interested reader to the recent survey~\cite{IM24}.

In the standard complex model without inhibitors, by indexing rows and columns by tests and items, respectively, a $t \times n$ $(d, r; z]$-disjunct matrix serves as the incidence matrix of a pooling design that allows one to decide whether each complex (a nonempty set of items) is defective in time linear in $t$, assuming at most $d$ defectives, complex size at most $r$, and at most $\lfloor  \frac{z - 1}{2} \rfloor$ erroneous outcomes~\cite{Mac97,DH06,CFH08}.

As described in the previous section, it is desirable for disjunct matrices to have large rates.
Hence, it is of central interest in group testing to understand the asymptotic behavior of the maximum possible number
\begin{equation*}
  N(t, d, r; z] \coloneqq \max \left\{ n : \begin{array}{l} \text{there is a $t \times n$} \\ \text{$(d, r; z]$-disjunct matrix} \end{array} \right\}
\end{equation*}
of columns of a $(d, r; z]$-disjunct matrix with exactly $t$ rows and the corresponding maximum achievable rate
\begin{equation*}
  R(t, d, r; z] \coloneqq \frac{\log_2 N(t, d, r; z]}{t}
\end{equation*}
for given positive integers $t$, $d$, $r$, and $z$.

The asymptotic behavior of $R(t, d, r; z]$ has been studied extensively~\cite{DR82,DRR89,Rus94,DFFT95,Eng96,Fur96,SWZ00,SW04,DSW04,Wei06,PR11,Bsh15,AB16,GRV20,BCE21,TF22,RV23}.
The best known asymptotic upper and lower bounds provided in~\cite[Th.~2.19, Th.~3.9]{SW04} state that, for any positive integer $z \leq \frac{c_0}{d^r} t$, we have
\begin{equation}
  \label{eq:bounds_disjunct}
  \frac{c_1 + o(1)}{d^{r + 1}} \leq R(t, d, r; z] \leq \frac{(c_2 + o(1)) \log_2 d}{d^{r + 1}}
\end{equation}
as $t \to \infty$, where $c_0$, $c_1$, and $c_2$ are positive constants independent of $t$, $d$, and $z$.
This implies that $t = \Theta(\log_2 n)$ tests suffice to identify all defectives under the standard complex model without inhibitors, even with up to $\lfloor \frac{z - 1}{2} \rfloor$ errors.

In contrast to the well-developed theory of conventional disjunct matrices, the asymptotic behavior of their inclusive variant is much less understood.
For disjoint sets $A, B \subseteq [n]$, define
\begin{equation*}
  Y_T(A, B) \coloneqq \left| \left( \bigcap_{j \in B} C_{j} \right) \cap \left( \bigcup_{j \in A} C_{j} \right) \right|
\end{equation*}
to be the number of rows in which at least one of the $|A|$ entries in the columns indexed by the elements of $A$ is $1$ and all the $|B|$ entries in the columns indexed by the elements of $B$ are $1$.
For a positive integer $r$ and nonnegative integers $h$ and $y$, the binary matrix $T$ is \emph{$(h, r; y]$-inclusive} if for any pair of disjoint sets $A, B \subseteq [n]$ with $|A| = h$ and $|B| = r$, $Y_T(A, B) \leq y$.
For positive integers $d$, $r$, and $x$, and a nonnegative integer $h$, we say that $T$ is \emph{$(d, h, r; x]$-inclusively disjunct} if there exist integers $z > 0$ and $y \geq 0$ with $z - y \geq x$ such that $T$ is simultaneously $(d, r; z]$-disjunct and $(h, r; y]$-inclusive.
It was shown in~\cite{CCF10} that a $t \times n$ $(d, h, r; x]$-inclusively disjunct matrix enables efficient identification of all defectives under the general inhibitor complex model with at most $d$ defectives, at most $h$ inhibitors, complex size at most $r$, and at most $\lfloor \frac{x - 1}{2} \rfloor$ erroneous outcomes not introduced by inhibitors.
The inference algorithm of~\cite{CCF10} decides whether each complex is defective in time linear in $t$, improving substantially over the approach of~\cite{CCH11}, which uses a $t\times n$ $(d + h, r; z]$-disjunct matrix and requires time $O(t n^h)$.

Define
\begin{equation*}
  N(t, d, h, r; x] \coloneqq \max \left\{ n : \begin{array}{l} \text{there is a $t \times n$ $(d, h, r; x]$} \\ \text{-inclusively disjunct matrix} \end{array} \right\}.
\end{equation*}
As with conventional disjunct matrices, the maximum achievable rate
\begin{equation*}
  R(t, d, h, r; x] \coloneqq \frac{\log_2 N(t, d, h, r; x]}{t}
\end{equation*}
is also a key parameter governing the efficiency of the inclusive variant.
The following lemma from~\cite{CCF10} yields an upper bound on $R(t, d, h, r; x]$.

\begin{lem}[{\cite[Lem.~5.1]{CCF10}}]
  Every $(d, h, r; x]$-inclusively disjunct matrix is $(d + h, r; x]$-disjunct.
\end{lem}

Combining this lemma with \eqref{eq:bounds_disjunct}, we have
\begin{align}
  \notag
  R(t, d, h, r; x]
  &\leq R(t, d + h, r; x] \\
  \label{eq:upper_bound_inclusively_disjunct}
  &\leq \frac{(c_2 + o(1)) \log_2 (d + h)}{(d + h)^{r + 1}}
\end{align}
as $t \to \infty$ for $x \leq \frac{c_0}{(d + h)^r} t$, where $c_0$ and $c_2$ are positive constants independent of $t$, $d$, $h$, and $x$.

While the inference algorithm of~\cite{CCF10} is highly efficient, most known inclusively disjunct matrices require a large number of rows.
For $r = 1$, special set systems first constructed as standard disjunct matrices in~\cite{HS87} were shown to be inclusively disjunct in~\cite[Lem.~3.3, Th.~3.4]{CCF10}, yielding the best known lower bound
\begin{equation*}
    R(t, d, h, 1; x] \geq \frac{\log_2 3}{16 (d + h + x - 1)^2} - \frac{\log_2 3 - 1}{t}.
\end{equation*}
If the error-tolerance parameter $x$ is fixed, then this lower bound implies
\begin{equation*}
  R(t, d, h, 1; x] \geq \frac{c + o(1)}{(d + h)^2}
\end{equation*}
for a positive constant $c$ independent of $t$, $d$, $h$, and $x$, which matches the upper bound~\eqref{eq:upper_bound_inclusively_disjunct} up to a logarithmic factor.
However, if $x$ is an increasing function of $t$, then the lower bound asymptotically vanishes.

Unfortunately, for general $r \geq 1$, most constructions of $(d, r; z]$-disjunct matrices do not guarantee inclusiveness.
In fact, although \cite{CCF10} pointed out that several constructions of $(d, r; z]$-disjunct matrices generate $(d, h, r; x]$-inclusively disjunct matrices, their outputs have asymptotically vanishing rates even when $x$ is a constant.
The difficulty is that strengthening disjunctness tends to increase the overlaps that inclusiveness is designed to limit; hence, constructions must balance these competing constraints.
This raises the following natural questions.
For general $r \geq 1$, can $(d, h, r; x]$-inclusively disjunct matrices achieve asymptotically positive rates?
If so, is there an efficient construction of such matrices?
In the next section, we resolve them affirmatively.

\section{Asymptotic lower bound and construction}
\label{sec:lower_bound_randomized_construction}
In this section, we establish the first nontrivial lower bound on $R(t, d, h, r; x]$ and give an efficient randomized construction of $(d, h, r; x]$-inclusively disjunct matrices that achieves this bound.
We prove the following theorem via the probabilistic method.

\begin{thm}
  \label{thm:lower_bound_inclusively_disjunct}
  For any fixed positive integer $r$, there exist positive constants $c_0$ and $c$ depending only on $r$ such that the following holds.
  For any fixed integers $d \geq 1$ and $h \geq 0$, and for any positive integers $t$ and $x \leq \frac{c_0}{(\max\{ d, h \})^r} t$,
  \begin{equation*}
    R(t, d, h, r; x] \geq \frac{c + o(1)}{(\max \{ d, h \})^{r + 1}},
  \end{equation*}
  where $o(1)$ is a function of $t$ satisfying $o(1) \to 0$ as $t \to \infty$.
\end{thm}

This asymptotic lower bound matches the upper bound \eqref{eq:upper_bound_inclusively_disjunct} up to a logarithmic factor.
Moreover, our guarantee allows the error-tolerance parameter $x$ to grow linearly with $t$, matching the linear-in-$t$ regime for the classical parameter $z$ in~\eqref{eq:bounds_disjunct} up to a constant factor.
Hence, our bound is a natural extension of the asymptotic lower bound on $R(t, d, r; z]$ shown in \eqref{eq:bounds_disjunct}.

To prove Theorem~\ref{thm:lower_bound_inclusively_disjunct}, we use two standard probabilistic tools: \emph{Markov's inequality} and the \emph{Chernoff bounds}~\cite{MU17}.

\begin{lem}[Markov's inequality]
  Let $X$ be a nonnegative random variable.
  Then, for any $a > 0$,
  \begin{equation*}
    \Pr[X \geq a] \leq \frac{\E[X]}{a}.
  \end{equation*}
\end{lem}

\begin{lem}[Chernoff bounds]
  Let $X_1,\allowbreak X_2,\allowbreak \dots,\allowbreak X_n$ be mutually independent random variables taking values in $\{ 0, 1 \}$, and define $X \coloneqq X_1 + X_2 + \dots + X_n$.
  Then, the following hold.
  \begin{enumerate}
    \item For any $\delta \in (0, 1)$,
    \begin{equation*}
      \Pr[X \leq (1 - \delta) \E[X]] \leq \exp\left( - \frac{\delta^2 \cdot \E[X]}{2} \right).
    \end{equation*}
    \item For any $\delta > 0$,
    \begin{equation*}
      \Pr[X \geq (1 + \delta) \E[X]] \leq \exp\left( - \frac{\delta^2 \cdot \E[X]}{2 + \delta} \right).
    \end{equation*}
  \end{enumerate}
\end{lem}

For convenience, we introduce the notion of violated sets.
Let $z > 0$, $y \geq 0$, and $x > 0$ be integers with $z - y \geq x$.
For a $t \times n$ binary matrix $T$ over $\{ 0, 1 \}$, we call $B \subseteq [n]$ \emph{$(d, h, r; z, y]$-violated} if $B$ is an $r$-set such that there exists a $d$-subset $A \subseteq [n] \setminus B$ satisfying $Z_T(A, B) \leq z - 1$ or there exists an $h$-subset $A' \subseteq [n] \setminus B$ satisfying $Y_T(A', B) \geq y + 1$.
If $T$ has no $(d, h, r; z, y]$-violated subset of $[n]$, then $T$ is simultaneously $(d, r; z]$-disjunct and $(h, r; y]$-inclusive, and hence it is $(d, h, r; x]$-inclusively disjunct because $z - y \geq x$.

We now prove Theorem~\ref{thm:lower_bound_inclusively_disjunct} using a powerful technique called the \emph{alteration method}~\cite{AS16} in probabilistic combinatorics, whereby we modify a random object to obtain the desired structure.
In the construction below, we first generate a random binary matrix with some extra columns.
We then select one element from each violated set and delete all columns indexed by them, yielding a desired inclusively disjunct matrix.

\begin{IEEEproof}[Proof of Theorem~\ref{thm:lower_bound_inclusively_disjunct}]
  Fix absolute constants $\varepsilon \in (0, 1)$, $\delta \in (0, 2^{\varepsilon} - 1)$, and $\sigma > 0$.
  Set $D \coloneqq \max \{ d, h \}$.
  Since any $(D, D, r; x]$-inclusively disjunct matrix is also $(d, h, r; x]$-inclusively disjunct, it suffices to construct a $(D, D, r; x]$-inclusively disjunct matrix.

  Let $n$ be a positive integer to be determined later.
  Set
  \begin{equation*}
    p \coloneqq 1 - 2^{- \frac{1 - \varepsilon}{D}}.
  \end{equation*}
  Since $\exp(-u) \leq 1 - u + \frac{u^2}{2}$ for any $u \geq 0$, we obtain
  \begin{align*}
    p
    &= 1 - \exp \left( - \frac{(1 - \varepsilon) \ln 2}{D} \right) \\
    &\geq 1 - \left( 1 - \frac{(1 - \varepsilon) \ln 2}{D} + \frac{((1 - \varepsilon) \ln 2)^2}{2 D^2} \right) \\
    &\geq \frac{(1 - \varepsilon) \ln 2}{2 D}.
  \end{align*}
  Let $T$ be a $t \times (1 + \sigma) n$ random matrix over $\{ 0, 1 \}$ whose entries independently take the value $1$ with probability $p$ and $0$ with probability $1 - p$.

  By the linearity of expectation, for any pair of disjoint sets $A, B \subseteq [(1 + \sigma) n]$ with $|A| = D$ and $|B| = r$, we have
  \begin{equation*}
    \E[Z_T(A, B)] = p^r (1 - p)^D \cdot t
  \end{equation*}
  and
  \begin{equation*}
    \E[Y_T(A, B)] = p^r (1 - (1 - p)^D) \cdot t.
  \end{equation*}
  Set
  \begin{equation*}
    z \coloneqq \lfloor (1 - \delta) p^r (1 - p)^D \cdot t \rfloor + 1
  \end{equation*}
  and
  \begin{equation*}
    y \coloneqq \lceil (1 + \delta) p^r (1 - (1 - p)^D) \cdot t \rceil - 1.
  \end{equation*}
  Then, we have
  \begin{align*}
    z - y
    &\geq p^r (2 (1 - p)^D - 1 - \delta) \cdot t \\
    &\geq \left( \frac{(1 - \varepsilon) \ln 2}{2 D} \right)^r (2^{\varepsilon} - 1 - \delta) \cdot t.
  \end{align*}
  By setting
  \begin{equation*}
    c_0 \coloneqq \left( \frac{(1 - \varepsilon) \ln 2}{2} \right)^r (2^{\varepsilon} - 1 - \delta)
  \end{equation*}
  and choosing a positive integer $x$ satisfying $x \leq \frac{c_0}{D^r} t$, we obtain $z - y \geq x$.

  By the Chernoff bounds, we have
  \begin{align*}
    &\Pr[Z_T(A, B) \leq z - 1] \\
    \leq{}& \Pr[Z_T(A, B) \leq (1 - \delta) \cdot \E[Z_T(A, B)]] \\
    \leq{}& \exp\left( - \frac{\delta^2 \cdot \E[Z_T(A, B)]}{2} \right) \\
    ={}& \exp\left( - \frac{\delta^2 p^r (1 - p)^D}{2} t \right)
  \end{align*}
  and
  \begin{align*}
    &\Pr[Y_T(A, B) \geq y + 1] \\
    \leq{}& \Pr[Y_T(A, B) \geq (1 + \delta) \cdot \E[Y_T(A, B)]] \\
    \leq{}& \exp\left( - \frac{\delta^2 \cdot \E[Y_T(A, B)]}{2 + \delta} \right) \\
    ={}& \exp\left( - \frac{\delta^2 p^r (1 - (1 - p)^D)}{2 + \delta} t \right).
  \end{align*}

  Recall that an $r$-subset $B \subseteq [(1 + \sigma) n]$ is $(D, D, r; z, y]$-violated if and only if there exists a $D$-subset $A \subseteq [(1 + \sigma) n] \setminus B$ satisfying either $Z_T(A, B) \leq z - 1$ or $Y_T(A, B) \geq y + 1$.
  Define $X$ as the number of $(D, D, r; z, y]$-violated subsets of $[(1 + \sigma) n]$.
  Then, by selecting one element from each $(D, D, r; z, y]$-violated subset of $[(1 + \sigma) n]$ and deleting all columns indexed by them, we obtain a $t \times n'$ $(D, D, r; x]$-inclusively disjunct matrix $T'$, where $n'$ is an integer satisfying $(1 + \sigma) n - X \leq n' \leq (1 + \sigma) n$.

  By the linearity of expectation and the union bound, we have
  \begin{align}
    \notag
    \E[X]
    &= \sum_{\substack{B \subseteq [(1 + \sigma) n],\\|B| = r}} \Pr[\text{$B$ is $(D, D, r; z, y]$-violated}] \\
    \notag
    &\leq \sum_{\substack{B \subseteq [(1 + \sigma) n],\\|B| = r}} \sum_{\substack{A \subseteq [(1 + \sigma) n] \setminus B,\\|A| = D}} (\Pr[Z_T(A, B) \leq z - 1] \\
    \label{eq:pessimistic_estimator}
    &\quad+ \Pr[Y_T(A, B) \geq y + 1]) \\
    \notag
    &< 2 (1 + \sigma)^{D + r} n^{D + r} \exp\left( - \frac{\delta^2 p^r (1 - (1 - p)^D)}{2 + \delta} t \right).
  \end{align}
  By Markov's inequality,
  \begin{align*}
    &\Pr[X \geq \sigma n] \\
    \leq{}& \frac{\E[X]}{\sigma n} \\
    <{}& \frac{2 (1 + \sigma)^{D + r} n^{D + r - 1}}{\sigma} \exp\left( - \frac{\delta^2 p^r (1 - (1 - p)^D)}{2 + \delta} t \right).
  \end{align*}
  Hence, the number $X$ of $(D, D, r; z, y]$-violated subsets of $[(1 + \sigma) n]$ is less than $\sigma n$ with positive probability whenever
  \begin{equation}
    \label{eq:sufficient_condition}
    \frac{2 (1 + \sigma)^{D + r} n^{D + r - 1}}{\sigma} \exp\left( - \frac{\delta^2 p^r (1 - (1 - p)^D)}{2 + \delta} t \right) \leq 1.
  \end{equation}
  Thus, $T'$ is a $t \times n'$ $(D, D, r; x]$-inclusively disjunct matrix such that $n + 1 \leq n' \leq (1 + \sigma) n$ with positive probability whenever $n$ satisfies \eqref{eq:sufficient_condition}.

  Fix
  \begin{equation*}
    n = \left\lfloor \left( \frac{\sigma}{2 (1 + \sigma)^{D + r}} \right)^{\frac{1}{D + r - 1}} \exp\left( \frac{\delta^2 p^r (1 - (1 - p)^D)}{(2 + \delta) (D + r - 1)} t \right) \right\rfloor
  \end{equation*}
  so that \eqref{eq:sufficient_condition} holds.
  Then, there exists a $t \times n'$ $(D, D, r; x]$-inclusively disjunct matrix such that $n + 1 \leq n' \leq (1 + \sigma) n$.
  This implies
  \begin{align*}
    R(t, d, h, r; x]
    &\geq R(t, D, D, r; x] \\
    &\geq \frac{\log_2 n'}{t} \\
    &\geq \frac{\log_2 (n + 1)}{t} \\
    &\geq \frac{\log_2 \sigma - 1 - (D + r) \log_2 (1 + \sigma)}{(D + r - 1) \cdot t} \\
    &\quad + \frac{\delta^2 p^r (1 - (1 - p)^D)}{(2 + \delta) (D + r - 1) \ln 2} \\
    &\geq \frac{(1 + o(1)) \delta^2 (1 - \varepsilon)^r (\ln 2)^{r - 1} (1 - 2^{\varepsilon - 1})}{2^r (2 + \delta) (D + r - 1) D^r} \\
    &\geq \frac{(1 + o(1)) \delta^2 (1 - \varepsilon)^r (\ln 2)^{r - 1} (1 - 2^{\varepsilon - 1})}{2^r (2 + \delta) r D^{r + 1}} \\
    &= \frac{c + o(1)}{D^{r + 1}}
  \end{align*}
  as $t \to \infty$, where $c$ is a positive constant defined by
  \begin{equation*}
    c \coloneqq \frac{\delta^2 (1 - \varepsilon)^r (\ln 2)^{r - 1} (1 - 2^{\varepsilon - 1})}{2^r (2 + \delta) r}.
  \end{equation*}
  The proof is complete.
\end{IEEEproof}

The above proof yields a simple Monte Carlo construction of inclusively disjunct matrices.
This construction is efficient in the sense that it runs in polynomial time $O(t n^{\max \{ d, h \} + r})$ in the matrix size $t n$ for fixed $d$, $h$, and $r$.

In the construction, we take $n$ satisfying \eqref{eq:sufficient_condition} so that $\Pr[X \geq \sigma n] < 1$.
However, this results in the number of columns becoming too small with a certain probability.
To reduce this failure probability, for example, one should take $n$ satisfying
\begin{align}
  \notag
  &\frac{2 (1 + \sigma)^{\max \{ d, h \} + r} n^{\max \{ d, h \} + r - 1}}{\sigma} \\
  \label{eq:high_probability_n}
  &\quad \cdot \exp\left( - \frac{\delta^2 p^r (1 - (1 - p)^{\max \{ d, h \}})}{2 + \delta} t \right) \leq \frac{1}{t^s}
\end{align}
instead of \eqref{eq:sufficient_condition}, where $s$ is any fixed positive constant.
In this case, since the failure probability $\Pr[X \geq \sigma n]$ is less than $\frac{1}{t^s} = o(1)$, the Monte Carlo construction succeeds with probability $1 - o(1)$.
Moreover, the resulting matrices achieve the same asymptotic rate guarantee as in the proof.
The high-probability version of the construction is shown in Fig.~\ref{alg:Monte_Carlo_construction}.

\begin{figure*}[!t]
  \noindent\rule{\textwidth}{0.4pt}\par
  \vspace{4pt}
  \centering
  \begin{algorithmic}[1]
    \REQUIRE absolute constants $\varepsilon \in (0, 1)$, $\delta \in (0, 2^{\varepsilon} - 1)$, $\sigma > 0$, and $s > 0$, positive integers $t$, $d$, and $r$, a nonnegative integer $h$, and a positive integer $x$ satisfying $x \leq \frac{c_0}{(\max \{ d, h \})^r} t$, where $c_0$ is the positive constant given in the proof of Theorem~\ref{thm:lower_bound_inclusively_disjunct}
    \ENSURE a $t \times n$ $(d, h, r; x]$-inclusively disjunct matrix with $\frac{\log_2 n}{t} \geq \frac{c + o(1)}{(\max \{ d, h \})^{r + 1}}$, where $c$ is the positive constant given in the proof of Theorem~\ref{thm:lower_bound_inclusively_disjunct}
    \STATE{$D \gets \max \{ d, h \}$ and set $p$, $z$, and $y$ as in the proof of Theorem~\ref{thm:lower_bound_inclusively_disjunct}}
    \STATE{Let $n$ be the maximum positive integer satisfying \eqref{eq:high_probability_n}}
    \STATE{Take a $t \times (1 + \sigma) n$ matrix $T = (t_{ij})$, where each entry $t_{ij}$ is independently set to $1$ with probability $p$ and $0$ with probability $1 - p$}
    \STATE{If the number of $(D, D, r; z, y]$-violated subsets of $[(1 + \sigma) n]$ is at least $\sigma n$, then return FAIL}
    \STATE{Select one element from each $(D, D, r; z, y]$-violated subset of $[(1 + \sigma) n]$ and delete all columns indexed by them from $T$}
    \RETURN $T$
  \end{algorithmic}
  \vspace{-4pt}
  \par\noindent\rule{\textwidth}{0.4pt}
  \caption{A polynomial-time Monte Carlo construction that succeeds with probability $1 - o(1)$.}
  \label{alg:Monte_Carlo_construction}
\end{figure*}

Finally, we note that the same lower bound can also be derived by employing another powerful tool in probabilistic combinatorics called the \emph{Lov\'{a}sz local lemma}~\cite{EL73,DSW04}.
Using the algorithmic Lov\'{a}sz local lemma~\cite{MT10}, one obtains an expected $O(t n^{\max \{ d, h \} + r + 1})$-time Las Vegas construction of inclusively disjunct matrices whose rates achieve the lower bound.

\section{Derandomization}
\label{sec:derandomized_construction}
Here, we derandomize the Monte Carlo construction in the proof of Theorem~\ref{thm:lower_bound_inclusively_disjunct}.

Our derandomization is based on a standard technique called the \emph{method of conditional expectations}~\cite{MR02,MU17}, also known as the \emph{method of conditional probabilities}~\cite{AS16}, which is widely used to obtain efficient deterministic constructions of combinatorial objects, including (hyper)graph colorings~\cite{AS16,MR02}, disjunct matrices~\cite{PR11}, and perfect hash families~\cite{Col08,Dou19,DC20}.
Recall that, in the proof of Theorem~\ref{thm:lower_bound_inclusively_disjunct}, we defined $X$ as the number of $(D, D, r; z, y]$-violated subsets of $[(1 + \sigma) n]$ for the $t \times (1 + \sigma) n$ random matrix $T$ and chose $n$ so that $\E[X] < \sigma n$.
In our derandomization process, we greedily fix the entries of $T$, aiming to keep the conditional expectation of $X$ below $\sigma n$.
Since the resulting deterministic matrix has fewer than $\sigma n$ $(D, D, r; z, y]$-violated subsets, deleting some columns as in the proof of Theorem~\ref{thm:lower_bound_inclusively_disjunct} yields the desired matrix.

The method of conditional expectations is typically used to derandomize algorithms based on rudimentary probabilistic approaches~\cite{AS16,MR02,Col08,PR11,Dou19,DC20}, whereas our Monte Carlo construction is based on the alteration method, which introduces an additional deletion step.
In many cases, the deletion step removes far fewer than $\sigma n$ columns, yielding an output matrix with a noticeably larger number of columns than the worst-case guarantee; this improves the rate only in lower-order terms, but can be meaningful for practical finite regimes.

The primary obstacle to an efficient implementation is that the conditional expectation of $X$ is difficult to compute directly at each iteration step.
To overcome this, we maintain an efficiently computable upper bound on this conditional expectation, obtained by applying the union bound as in~\eqref{eq:pessimistic_estimator}, and we choose each entry so as to keep this upper bound below $\sigma n$ throughout the process.
This approach is known as the \emph{method of pessimistic estimators}~\cite{Rag88,AS16,WX08}.

We show the following theorem.

\begin{thm}
  \label{thm:derandomized_construction}
  For any fixed positive integer $r$, there exist positive constants $c_0$ and $c$ depending only on $r$ such that the following holds.
  For any fixed integers $d \geq 1$ and $h \geq 0$, and for any positive integers $t$ and $x \leq \frac{c_0}{(\max\{ d, h \})^r} t$, there exists a deterministic algorithm that constructs a $t \times n$ $(d, h, r; x]$-inclusively disjunct matrix in time $O(t n^{\max \{ d, h \} + r + 1})$ for some positive integer $n$ satisfying
  \begin{equation*}
    \frac{\log_2 n}{t} \geq \frac{c + o(1)}{(\max \{ d, h \})^{r + 1}},
  \end{equation*}
  where $o(1)$ is a function of $t$ satisfying $o(1) \to 0$ as $t \to \infty$.
\end{thm}

\begin{IEEEproof}
  Fix absolute constants $\varepsilon \in (0, 1)$, $\delta \in (0, 2^{\varepsilon} - 1)$, and $\sigma > 0$ and set $p$, $z$, $y$, $n$, $c_0$, and $c$ as in the proof of Theorem~\ref{thm:lower_bound_inclusively_disjunct}.
  Set $D \coloneqq \max \{ d, h \}$.
  We now construct a $t \times (1 + \sigma) n$ matrix over $\{ 0, 1 \}$ by fixing its entries one by one.

  Suppose that we have already determined all entries in the first $i - 1$ rows and $j - 1$ entries in the $i$th row, and are trying to determine the value of the $j$th entry $t_{ij}$ in the $i$th row.
  We refer to this as the \emph{$(i, j)$-step}.
  Let $T$ be a $t \times (1 + \sigma) n$ partially random matrix in which all determined entries are fixed and the others independently take the value $1$ with probability $p$ and $0$ with probability $1 - p$.
  Write $X$ for the number of $(D, D, r; z, y]$-violated subsets of $[(1 + \sigma) n]$ for $T$.
  Define
  \begin{align*}
    P_{\mathrm{prev}} &\coloneqq \sum_{\substack{B \subseteq [(1 + \sigma) n],\\|B| = r}} \sum_{\substack{A \subseteq [(1 + \sigma) n] \setminus B,\\|A| = D}} (\Pr[Z_T(A, B) \leq z - 1] \\
    &\quad + \Pr[Y_T(A, B) \geq y + 1]).
  \end{align*}
  By the linearity of expectation and the union bound, we have $\E[X] \leq P_{\mathrm{prev}}$.
  For $k \in \{ 0, 1 \}$, define
  \begin{align*}
    P_{A, B}(k) &\coloneqq \Pr[Z_T(A, B) \leq z - 1 \mid t_{ij} = k] \\
    &\quad + \Pr[Y_T(A, B) \geq y + 1 \mid t_{ij} = k]
  \end{align*}
  and
  \begin{equation*}
    P_{\mathrm{curr}}(k) \coloneqq \sum_{\substack{B \subseteq [(1 + \sigma) n],\\|B| = r}} \sum_{\substack{A \subseteq [(1 + \sigma) n] \setminus B,\\|A| = D}} P_{A, B}(k).
  \end{equation*}
  Then, $\E[X \mid t_{ij} = k] \leq P_{\mathrm{curr}}(k)$.
  By the law of total probability, we have
  \begin{align*}
    P_{\mathrm{prev}}
    &= \sum_{\substack{B \subseteq [(1 + \sigma) n],\\|B| = r}} \sum_{\substack{A \subseteq [(1 + \sigma) n] \setminus B,\\|A| = D}} ((1 - p) P_{A, B}(0) \\
    &\quad + p P_{A, B}(1)) \\
    &= (1 - p) \cdot P_{\mathrm{curr}}(0) + p \cdot P_{\mathrm{curr}}(1) \\
    &\geq \min_{k \in \{ 0, 1 \}} P_{\mathrm{curr}}(k).
  \end{align*}
  Hence, if
  \begin{equation}
    \label{eq:assumption_induction}
    P_{\mathrm{prev}} < \sigma n,
  \end{equation}
  then we can take $k \in \{ 0, 1 \}$ such that $P_{\mathrm{curr}}(k) \leq P_{\mathrm{prev}}$ and thus
  \begin{align*}
    \E[X \mid t_{ij} = k]
    &\leq P_{\mathrm{curr}}(k) \\
    &\leq P_{\mathrm{prev}} \\
    &< \sigma n.
  \end{align*}
  As shown in the proof of Theorem~\ref{thm:lower_bound_inclusively_disjunct}, \eqref{eq:assumption_induction} holds in the $(1, 1)$-step.
  Therefore, by induction, we can determine all entries greedily so that the resulting deterministic matrix has less than $\sigma n$ $(D, D, r; z, y]$-violated subsets.
  From the proof of Theorem~\ref{thm:lower_bound_inclusively_disjunct}, by selecting one element from each $(D, D, r; z, y]$-violated subset and deleting all columns indexed by them, we obtain a desired matrix.

  It remains to analyze the running time.
  Define $q_1 \coloneqq p^r (1 - p)^D$ and $q_2 \coloneqq p^r (1 - (1 - p)^D)$.
  To reduce the running time, we precalculate
  \begin{equation*}
    s_1(z_0, i) \coloneqq \sum_{l = 0}^{z - z_0}  \binom{t - i}{l} q_1^l (1 - q_1)^{t - i - l}
  \end{equation*}
  and
  \begin{equation*}
    s_2(y_0, i) \coloneqq \sum_{l = 0}^{y - y_0}  \binom{t - i}{l} q_2^l (1 - q_2)^{t - i - l}
  \end{equation*}
  for each $z_0 \in [z]$, $y_0 \in \{ 0, 1, \dots, y \}$, and $i \in \{ 0, 1, \dots, t \}$ in time polynomial in $t$ and store the values in a lookup table.
  For convenience, we define $s_1(z_0, i) = 0$ and $s_2(y_0, i) = 0$ for any $z_0 > z$ and $y_0 > y$.

  As before, consider the $(i, j)$-step.
  Suppose that we have already calculated the value of $P_{\mathrm{prev}}$.
  Write $T_{\mathrm{fixed}}$ for the $(i - 1) \times (1 + \sigma) n$ deterministic matrix consisting of the first $i - 1$ rows of $T$ and $T_{\mathrm{random}}$ for the $(t - i + 1) \times (1 + \sigma) n$ random matrix consisting of the remaining rows.
  Let $A$ and $B$ be disjoint subsets of $[(1 + \sigma) n]$ such that $|A| = D$ and $|B| = r$.
  Since all entries in $T_{\mathrm{fixed}}$ have already been determined, both $z_{A, B} \coloneqq Z_{T_{\mathrm{fixed}}}(A, B)$ and $y_{A, B} \coloneqq Y_{T_{\mathrm{fixed}}}(A, B)$ are deterministic.
  Hence, for each $k \in \{ 0, 1 \}$, we have
  \begin{align*}
    P_{A, B}(k)
    &= \Pr[Z_T(A, B) \leq z - 1 \mid t_{ij} = k] \\
    &\quad + \Pr[Y_T(A, B) \geq y + 1 \mid t_{ij} = k] \\
    &= \Pr[Z_{T_{\mathrm{random}}}(A, B) \leq z - z_{A, B} - 1 \mid t_{ij} = k] \\
    &\quad + 1 - \Pr[Y_{T_{\mathrm{random}}}(A, B) \leq y - y_{A, B} \mid t_{ij} = k].
  \end{align*}
  Define
  \begin{equation}
    \label{eq:omega}
    \omega_{A, B} \coloneqq \Pr \left[ i \in \left( \bigcap_{l \in B} C_l \right) \setminus \left( \bigcup_{l \in A} C_l \right) \mathrel{}\middle|\mathrel{} t_{ij} = 0 \right]
  \end{equation}
  and
  \begin{equation}
    \label{eq:psi}
    \psi_{A, B} \coloneqq \Pr \left[ i \in \left( \bigcap_{l \in B} C_l \right) \cap \left( \bigcup_{l \in A} C_l \right) \mathrel{}\middle|\mathrel{} t_{ij} = 0 \right].
  \end{equation}
  Then, $\omega_{A, B}$ and $\psi_{A, B}$ can be computed in constant time (independent of $t$ and $n$) by inspecting only the $D + r$ entries indexed by $A \cup B$.
  By the law of total probability, we have
  \begin{align*}
    &\Pr[Z_{T_{\mathrm{random}}}(A, B) \leq z - z_{A, B} - 1 \mid t_{ij} = 0] \\
    ={}& \omega_{A, B} \cdot \sum_{l = 0}^{z - z_{A, B} - 2}  \binom{t - i}{l} q_1^l (1 - q_1)^{t - i - l} \\
    &+ (1 - \omega_{A, B}) \cdot \sum_{l = 0}^{z - z_{A, B} - 1}  \binom{t - i}{l} q_1^l (1 - q_1)^{t - i - l} \\
    ={}& \omega_{A, B} \cdot s_1(z_{A, B} + 2, i) + (1 - \omega_{A, B}) \cdot s_1(z_{A, B} + 1, i)
  \end{align*}
  and
  \begin{align*}
    &\Pr[Y_{T_{\mathrm{random}}}(A, B) \leq y - y_{A, B} \mid t_{ij} = 0] \\
    ={}& \psi_{A, B} \cdot \sum_{l = 0}^{y - y_{A, B} - 1} \binom{t - i}{l} q_2^l (1 - q_2)^{t - i - l} \\
    &+ (1 - \psi_{A, B}) \cdot \sum_{l = 0}^{y - y_{A, B}} \binom{t - i}{l} q_2^l (1 - q_2)^{t - i - l} \\
    ={}& \psi_{A, B} \cdot s_2(y_{A, B} + 1, i) + (1 - \psi_{A, B}) \cdot s_2(y_{A, B}, i).
  \end{align*}
  Therefore, we have
  \begin{align}
    \notag
    P_{A, B}(0)
    &= \omega_{A, B} \cdot s_1(z_{A, B} + 2, i) \\
    \notag
    &\quad + (1 - \omega_{A, B}) \cdot s_1(z_{A, B} + 1, i) \\
    \notag
    &\quad + 1 - \psi_{A, B} \cdot s_2(y_{A, B} + 1, i) \\
    \label{eq:P_AB_0}
    &\quad - (1 - \psi_{A, B}) \cdot s_2(y_{A, B}, i).
  \end{align}
  Since each $P_{A, B}(0)$ can be calculated in constant time (independent of $t$ and $n$) by random access to the precalculated lookup table, we can compute $P_{\mathrm{curr}}(0)$ in time $O(n^{D + r})$.
  If $P_{\mathrm{curr}}(0) \leq P_{\mathrm{prev}}$, then determine the value of $t_{ij}$ to be $0$; otherwise, determine it to be $1$.

  After completing each row $i$, we update all counters $z_{A, B}$ and $y_{A, B}$.
  This takes time $O(n^{D + r})$ per row because, for each pair of disjoint sets $A, B \subseteq [(1 + \sigma) n]$ with $|A| = D$ and $|B| = r$, the increment is determined by inspecting only the $D + r$ entries in that row indexed by $A \cup B$.
  This update cost is dominated by the $O(n^{D + r + 1})$ time spent to process the entries in the row.

  Therefore, the total running time is $O(t n^{D + r + 1})$.
  The proof is complete.
\end{IEEEproof}

The resulting deterministic construction is summarized in Fig.~\ref{alg:derandomized_construction}.

\begin{figure*}[!t]
  \noindent\rule{\textwidth}{0.4pt}\par
  \vspace{4pt}
  \centering
  \begin{algorithmic}[1]
    \REQUIRE absolute constants $\varepsilon \in (0, 1)$, $\delta \in (0, 2^{\varepsilon} - 1)$, and $\sigma > 0$, positive integers $t$, $d$, and $r$, a nonnegative integer $h$, and a positive integer $x$ satisfying $x \leq \frac{c_0}{(\max \{ d, h \})^r} t$, where $c_0$ is the positive constant given in the proof of Theorem~\ref{thm:lower_bound_inclusively_disjunct}
    \ENSURE a $t \times n$ $(d, h, r; x]$-inclusively disjunct matrix with $\frac{\log_2 n}{t} \geq \frac{c + o(1)}{(\max \{ d, h \})^{r + 1}}$, where $c$ is the positive constant given in the proof of Theorem~\ref{thm:lower_bound_inclusively_disjunct}
    \STATE{$D \gets \max \{ d, h \}$ and set $p$, $z$, $y$, and $n$ as in the proof of Theorem~\ref{thm:lower_bound_inclusively_disjunct}}
    \STATE{Calculate $s_1(z_0, i) = \sum_{l = 0}^{z - z_0}  \binom{t - i}{l} q_1^l (1 - q_1)^{t - i - l}$ and $s_2(y_0, i) = \sum_{l = 0}^{y - y_0}  \binom{t - i}{l} q_2^l (1 - q_2)^{t - i - l}$ for each $z_0 \in [z]$, $y_0 \in \{ 0, 1, \dots, y \}$, and $i \in \{ 0, 1, \dots, t \}$, where $q_1 = p^r (1 - p)^{D}$ and $q_2 = p^r (1 - (1 - p)^{D})$}
    \STATE{$P_{\mathrm{prev}} \gets \binom{(1 + \sigma) n}{r} \binom{(1 + \sigma) n - r}{D} (s_1(1, 0) + 1 - s_2(0, 0))$}
    \STATE{Take a $t \times (1 + \sigma) n$ matrix $T = (t_{ij})$ filled with the symbol $*$, which represents an undetermined value}
    \STATE{$z_{A, B} \gets 0$ and $y_{A, B} \gets 0$ for each pair of disjoint sets $A, B \subseteq [(1 + \sigma) n]$ with $|A| = D$ and $|B| = r$}
    \FOR{$i = 1$ to $t$}
      \FOR{each entry $t_{ij}$ in the $i$th row of $T$}
        \STATE{Calculate $\omega_{A, B}$ as in~\eqref{eq:omega}, $\psi_{A, B}$ as in~\eqref{eq:psi}, and $P_{A, B}(0)$ as in~\eqref{eq:P_AB_0} for each pair of disjoint sets $A, B \subseteq [(1 + \sigma) n]$ with $|A| = D$ and $|B| = r$}
        \STATE{$P_{\mathrm{curr}}(0) \gets \sum_{\substack{B \subseteq [(1 + \sigma) n],\\|B| = r}} \sum_{\substack{A \subseteq [(1 + \sigma) n] \setminus B,\\|A| = D}} P_{A, B}(0)$ and $P_{\mathrm{curr}}(1) \gets \frac{P_{\mathrm{prev}} - (1 - p) P_{\mathrm{curr}}(0)}{p}$}
        \STATE{Fix $t_{ij}$ to $0$ if $P_{\mathrm{curr}}(0) \leq P_{\mathrm{prev}}$; otherwise, fix it to $1$}
        \STATE{$P_{\mathrm{prev}} \gets P_{\mathrm{curr}}(t_{ij})$}
      \ENDFOR
      \STATE{For each pair of disjoint sets $A, B \subseteq [(1 + \sigma) n]$ with $|A| = D$ and $|B| = r$, increment $z_{A,B}$ and $y_{A,B}$ by $1$ if the $i$th row contributes to $Z_{T_{\mathrm{fixed}}}(A,B)$ and $Y_{T_{\mathrm{fixed}}}(A,B)$, respectively, where $T_{\mathrm{fixed}}$ is the $i \times (1 + \sigma) n$ matrix consisting of the first $i$ determined rows of $T$}
    \ENDFOR
    \STATE{Select one element from each $(D, D, r; z, y]$-violated subset of $[(1 + \sigma) n]$ and delete all columns indexed by them from $T$}
    \RETURN $T$
  \end{algorithmic}
  \vspace{-4pt}
  \par\noindent\rule{\textwidth}{0.4pt}
  \caption{A deterministic polynomial-time construction.}
  \label{alg:derandomized_construction}
\end{figure*}

\section{Concluding remarks}
\label{sec:concluding_remarks}
We established the first nontrivial asymptotic lower bound on the maximum achievable rate of inclusively disjunct matrices, matching the strongest known upper bound up to a logarithmic factor.
Our proof yields a simple randomized construction that achieves this bound, and we derandomized it to obtain a deterministic polynomial-time construction.
These results suggest that, even under the most adversarial inhibitors, large-scale error-tolerant group testing can be carried out with asymptotic efficiency broadly comparable to that of the classical case.
This provides further motivation for studying inclusively disjunct matrices in applications where inhibitor behavior is hard to predict.

While our constructions run in polynomial time in the matrix size $t n$, one may prefer faster constructions in practice.
For conventional disjunct matrices, linear-time constructions of $(d, 1; 1]$- and $(d, r; 1]$-disjunct matrices were provided in~\cite{PR11} and~\cite{Bsh15}, respectively.
It is natural to ask whether similar techniques can be adapted to the inclusive variant; however, the inclusiveness constraint obstructs a straightforward application, and the resulting matrices do not in general guarantee $(d, h, r; x]$-inclusively disjunctness.
In fact, we can deterministically construct inclusively disjunct matrices in linear time in the matrix size for $r = 1$ by extending the approach in~\cite{PR11} and~\cite{Bsh15}, yet for $r \geq 2$, the need to control all $r$-wise intersections simultaneously makes a comparable linear-time construction seem out of reach.
Progress in this direction would be an important step toward making inclusively disjunct matrices broadly usable in practice.

\IEEEtriggeratref{50}
\IEEEtriggercmd{\enlargethispage{-5.39in}}


\end{document}